\documentclass[12pt,epsf]{article}
\usepackage{graphicx}
\usepackage{epsfig}
\begin{document}

\def\a{\alpha}
\def\b{\beta}
\def\c{\varepsilon}
\def\d{\delta}
\def\e{\epsilon}
\def\f{\phi}
\def\g{\gamma}
\def\h{\theta}
\def\k{\kappa}
\def\l{\lambda}
\def\m{\mu}
\def\n{\nu}
\def\p{\psi}
\def\q{\partial}
\def\r{\rho}
\def\s{\sigma}
\def\t{\tau}
\def\u{\upsilon}
\def\v{\varphi}
\def\w{\omega}
\def\x{\xi}
\def\y{\eta}
\def\z{\zeta}
\def\D{\Delta}
\def\G{\Gamma}
\def\H{\Theta}
\def\L{\Lambda}
\def\F{\Phi}
\def\P{\Psi}
\def\S{\Sigma}

\def\o{\over}
\def\beq{\begin{eqnarray}}
\def\eeq{\end{eqnarray}}
\newcommand{\gsim}{ \mathop{}_{\textstyle \sim}^{\textstyle >} }
\newcommand{\lsim}{ \mathop{}_{\textstyle \sim}^{\textstyle <} }
\newcommand{\vev}[1]{ \left\langle {#1} \right\rangle }
\newcommand{\bra}[1]{ \langle {#1} | }
\newcommand{\ket}[1]{ | {#1} \rangle }
\newcommand{\EV}{ {\rm eV} }
\newcommand{\KEV}{ {\rm keV} }
\newcommand{\MEV}{ {\rm MeV} }
\newcommand{\GEV}{ {\rm GeV} }
\newcommand{\TEV}{ {\rm TeV} }
\def\diag{\mathop{\rm diag}\nolimits}
\def\Spin{\mathop{\rm Spin}}
\def\SO{\mathop{\rm SO}}
\def\O{\mathop{\rm O}}
\def\SU{\mathop{\rm SU}}
\def\U{\mathop{\rm U}}
\def\Sp{\mathop{\rm Sp}}
\def\SL{\mathop{\rm SL}}
\def\tr{\mathop{\rm tr}}

\def\IJMP{Int.~J.~Mod.~Phys. }
\def\MPL{Mod.~Phys.~Lett. }
\def\NP{Nucl.~Phys. }
\def\PL{Phys.~Lett. }
\def\PR{Phys.~Rev. }
\def\PRL{Phys.~Rev.~Lett. }
\def\PTP{Prog.~Theor.~Phys. }
\def\ZP{Z.~Phys. }

\newcommand{\bear}{\begin{array}}  
\newcommand {\eear}{\end{array}}
\newcommand{\la}{\left\langle}  
\newcommand{\ra}{\right\rangle}
\newcommand{\non}{\nonumber}  
\newcommand{\ds}{\displaystyle}
\newcommand{\red}{\textcolor{red}}
\newcommand{\mwino}{m_{\widetilde{W}^0}}
\def\ubl{U(1)$_{\rm B-L}$}
\def\REF#1{(\ref{#1})}
\def\lrf#1#2{ \left(\frac{#1}{#2}\right)}
\def\lrfp#1#2#3{ \left(\frac{#1}{#2} \right)^{#3}}
\def\OG#1{ {\cal O}(#1){\rm\,GeV}}


\baselineskip 0.7cm

\begin{titlepage}

\vskip 1.35cm
\begin{center}
{\large \bf
Considerations of Cosmic Acceleration}
\vskip 1.35cm
Paul H. Frampton\footnote{e-mail: frampton@physics.unc.edu}
\vskip 0.4cm
{\it Department of Physics and Astronomy, University of North Carolina,
Chapel Hill, NC 27599-3255}

\vskip 1.5cm

\abstract{ 
I discuss a solution to the dark energy
problem, which arises when the
visible universe is approximated by a black hole,
in a quasi-static asymptotically-flat approximation.
Using data, provided by WMAP7, I calculate the 
Schwarzschild radius $r_S$ and compare to the
measured physical radius of
the visible universe, bounded 
by the surface of last scatter. The ratio, $\epsilon(t_0)  = r/r_S$ is
found to be comparable to $\epsilon = 1$,
as allowed by the holographic principle.
The measurement of a shift parameter, $\sigma$, 
introduced by Bond, Efstathiou and Tegmark in 1997,
plays an important role in the accuracy of the calculation.
The approximation leads to a surprisingly small discrepancy,
presumably explicable by
the de Sitter, and expanding, nature of the actual universe.}

\end{center}
\end{titlepage}

\bigskip
\bigskip

\section{Introduction}

\bigskip

\noindent 
In fundamental theoretical physics, there was,
at the beginning of the twenty-first century,
an impossible seeming problem. The problem
is the dark energy in cosmology, comprising some
seventy percent of the universe.

\bigskip

\noindent
Another cosmological problem, closely entwined with the
dark energy problem, is the question well posed, 
now almost eight decades ago,
by Tolman\cite{Tolman} as to whether one can construct 
a consistent cyclic model, given the seemingly
contradictory constraint imposed
by the second law of thermodynamics.
The most developed solution of this Tolman conundrum is that
suggested in \cite{BF}.

\bigskip

\noindent 
The most important observational advance in cosmology since 
the early studies of cosmic expansion in the 1920s was the dramatic and,
at that time, surprising
discovery in the
waning years of the twentieth century that the expansion is accelerating.
This was first announced in February 1998, and it was based on the concordance of
two groups' data on Supernovae Type 1A \cite{Perlmutter, Reiss}

\bigskip

\noindent
Many subsequent experiments concerning
the Cosmic Microwave Background (CMB), Large Scale Structure (LSS)
and other measurements have all
confirmed the 1998 claim.
I therefore adopt the position that the accelerated expansion
rate is an observed fact.

\bigskip

\noindent Assuming general relativity, together
with the cosmological principle of homogeneity and isotropy,
the scale factor $a(t)$ in the FRW metric satisfies \cite{F,L}
the Friedmann-Lema\^{i}tre equation, with different energy
density components subsumed into $\rho$:

\begin{equation}
H(t)^2 = \left( \frac{\dot{a}}{a} \right)^2 = \left( \frac{8 \pi G}{3} \right) \rho. 
\label{FLequation}
\end{equation}

\bigskip

\noindent I normalize $a(t_0) = 1$ at the present, time $t=t_0$,
and $\rho$ is an energy density source which drives the
expansion of the universe. Two established contributions to $\rho$
are $\rho_m$ from matter (including dark matter) and $\rho_{\gamma}$
radiation, so that

\begin{equation}
\rho \supseteq \rho_{m} + \rho_{\gamma}
\label{rho}
\end{equation}

\noindent with  $\rho_{m}(t)  =  \rho_{m}(t_0) a(t)^{-3}$
and $\rho_{\gamma} (t)  = \rho_{\gamma}(t_0) a(t)^{-4}$.

\bigskip

\noindent For the observed accelerated expansion, a phenomenological 
approach is to add to the sources, in Eq.(\ref{FLequation}), a
dark energy term $\rho_{DE}(t)$ with 

\begin{equation}
\rho_{DE} (t) = \rho_{DE}(t_0) a(t)^{-3 (1 + \omega)},
\label{rhoDE}
\end{equation}

\bigskip
\bigskip
\noindent where $\omega = p/\rho$ is the equation of state. For the
case $\omega = -1$, as for a cosmological constant $\Lambda$,
and if one discards the matter and radiation terms which are smaller
than the dark energy term, one can easily
integrate the Friedmann-Lema\^{i}tre equation to find 

\begin{equation}
a(t) = a(t_0) ~ e^{ H t}
\label{CC}
\end{equation}

\bigskip

\noindent where $\sqrt{3} H= \sqrt{\Lambda} = \sqrt {8 \pi G \rho_{DE}}$.

\bigskip

\noindent By differentiation of Eq. (\ref{CC}) with respect
to time $p$ times, one obtains for the $p^{th}$
derivative

\begin{equation}
\frac{d^p}{dt^p} a(t) |_{t=0} = (H)^p
\label{jerk}
\end{equation}

\bigskip

\noindent Therefore, if $\Lambda$ is positive,
as in a De Sitter geometry, not only is the
acceleration ($p=2$) positive and non-zero,
but so are the jerk ($p=3$), the snap ($p=4$),
the crackle ($p=5$), the pop ($p=6$) and
all derivatives for $p \ge 7$.

\bigskip

\noindent The insertion of the dark energy term
Eq.~(\ref{rhoDE}) in Eq.~(\ref{FLequation})
works very well as a part of the $\Lambda CDM$
model. However, it is an {\it ad hoc} procedure
which gives no insight into what dark energy is.

\bigskip

\section{Dark Energy Problem}

\bigskip

\noindent  With this background, I shall now move
to a different explanation for the accelerated
expansion which obviates any dark energy, including
any need for a cosmological constant.

\bigskip

\noindent
The following considerations may initially appear to be trivial, circular,
tautological or some combination thereof. It appears, nevertheless,
by hindsight that the 1998 discovery of cosmic acceleration
could have been far less surprising theoretically had one
previously thought of the $\Lambda = 0$ universe as a black hole.

\bigskip

\noindent I now adopt this different approach,
with no dark energy, where instead the
central role is played by the assumption
of the holographic principle\cite{Hooft,Susskind} ( see also \cite{Bousso}) 
and by the overriding concept of entropy. 

\bigskip

\noindent 
The essential assumption is the aforementioned holographic principle,
by which I understand that all the information
about the universe is encoded on its
two-dimensional surface. What this implies is, however unlikely
it seems and however contrary to everyday experience that
the three-dimensional world I apparently observe
is somehow an illusion. This can
lead to a reinterpretation of the cosmic
acceleration, and possibly the most dramatic new insight into
gravity in over three centuries. 

\bigskip

\noindent
Consider the Schwarzschild
radius ($r_s$), and the physical radius ($R$), of the Sun ($\odot$).
They are $(r_s)_{\odot} = 3 km$ and $R_{\odot} = 800,000 km$.
Their ratio is $(\rho)_{\odot} \equiv (R/r_s)_{\odot} = 2.7 \times 10^5$.
One can readily check that for the Earth or for the Milky Way
the ratio $\rho = (R/r_s)$ is likewise much larger than one: $\rho >> 1$.
Such objects are nowhere close to being  black hole. Now 
consider the visible universe (VU) up to the surface of last scatter.
As we shall disucss later, with mass $M_{VU} \sim 5.5 \times 10^{23}
M_{\odot}$, it has $(r_s)_{VU}$ surprisingly close t $(r)_{VU} \sim 14 Gpc$,
even with the approximations of a quasi-static and
asymptotically-flat universe. The visible universe, within which we all live,
is well approximated by a black hole. The following addressing of the dark energy
problem follows from this observation.

\bigskip

\noindent
At the horizon, there is a PBH
temperature \cite{Parker, Bekenstein,Hawking}, $T_{\beta} $, which I can estimate as

\begin{equation}
T_{\beta} = \frac{\hbar}{  k_B }~\frac{H }{ 2 \pi} \sim 3 \times 10^{-30} K .
\label{T-beta}
\end{equation}

\bigskip

\noindent This temperature of the horizon information screen leads
to a concomitant  FDU acceleration \cite{Fulling,Davies,Unruh}
 $a_{Horizon}$, outward, of the horizon given
by the relationship

\bigskip

\begin{equation}
a_{Horizon} = \left( \frac{2 \pi c k_B T_{\beta}}{\hbar} \right) = c H \sim 10^{-9} \, m/s^2
\,.
\label{acceleration}
\end{equation}

\bigskip

\noindent When $T_{\beta}$ is used in Eq. (\ref{acceleration}),
I arrive at a cosmic acceleration which is essentially in
agreement with the observations\cite{Perlmutter, Reiss}.

\bigskip

\bigskip

\noindent
From this viewpoint, the dark energy is non-existent.
Instead there is a consequence of the second law
of thermodynamics, acting 
to create the appearance of a dark energy component
of the driving density on the right-hand-side of the
Friedman-Lema\^{i}tre equation, Eq.(\ref{FLequation}).

\bigskip

\noindent 
I have discussed a theory underlying the accelerated expansion
of the universe based on entropy. This approach 
provides a physical understanding of the acceleration phenomenon
which was lacking in the description as dark energy.

\bigskip

\noindent 
The entropy of the universe has received some recent attention~\cite{Lineweaver},
in  part because it relates to the feasibility of constructing a consistent cyclic model.
For example, the cyclic model in~\cite{BF}, assuming its internal consistency will indeed
be fully confirmed, provides the solution to a difficult entropy question
originally posed,  seventy-five years earlier, by Tolman~\cite{Tolman}.
The accelerated expansion rate is no longer surprising. It is the inevitable
consequence of information storage on the surface of
the visible universe.

\bigskip

\noindent
This solution of the dark energy problem not only
solves a cosmological problem but also casts a completely new light on the
nature of the gravitational force
\cite{Verlinde,Jacobson,Padmanaban,Ng}. Since the expansion of the universe,
including the acceleration thereof,
can only be a gravitational phenomenon, I arrive at the viewpoint that
gravity is a classical result of the second law of thermodynamics.
This means that gravity cannot be regarded as on the same footing
with the electroweak and strong interactions. Only if one assumes 
higher spatial dimensions, it can be argued\cite{Freund}
that non-gravitational interactions are  also be emergent.

\bigskip

\noindent
Although this can be the most radical change in gravity
theory for three centuries, it is worth emphasizing
that general relativity and its classical tests remain
unscathed, as does the prediction of gravitational waves.

\bigskip

\noindent
The result calls into question almost all of the work done
on quantum gravity, since the discovery
of quantum mechanics. For gravity, there is no longer necessity
for a graviton. In the case of string theory, the principal
motivation\cite{yoneya,ScherkSchwarz} for the 
profound and historical suggestion
by Scherk and Schwarz that string theory be reinterpreted,
not as a theory of the strong interaction, but instead as
a theory of the gravitational interaction,
came from the natural appearance of a massless graviton
in the closed string sector.

\bigskip

\noindent
This is not saying that string theory is dead.  What it is 
saying is, that string theory cannot be a theory of the
fundamental gravitational interaction, since there is
no fundamental gravitational interaction.

\bigskip

\noindent
The way this new insight emerged, as a solution of
the dark energy problem itself, was as a natural line
of thought, following the discovery of a cyclic model in
\cite{BF}, and the subsequent investigations
\cite{Entropy1,Entropy2,Entropy3,EFS1,EFS2} of the
entropy
of the universe, including a possible candidate
for dark matter\cite{Entropy3,FKTY}.
It is also strongly overlapping with the
discussions in \cite{Verlinde,Jacobson}.

\bigskip

\noindent
Another ramification, of such a solution of the dark energy,
problem is the status, fundamental versus emergent,
of the three spatial dimensions,
that we all observe every day. Because the solution
assumes the holographic principle\cite{Hooft,Susskind},
at least one spatial dimension appears as emergent.
Regarding the visible universe as a sphere,
with radius of about 14 Gpc, the emergent space
dimension is then, in spherical polar coordinates,
the radial coordinate, while the other two coordinates,
the polar and azimuthal angles, remain fundamental.
Physical intuition, related to the isotropy of space,
may suggest that, if one space dimension is emergent,
then so must be all three. This merits further investigation,
and may require a generalization of the
holographic principle in \cite{Hooft,Susskind}.
On the other hand, a fundamental time
coordinate is useful in dynamics.

\bigskip

\noindent
Of course, this present discussion
of cosmic acceleration, is
merely one small step towards the ultimate goal, of a
cyclic model, in which time never begins or ends.

\bigskip

\section{Holographic Principle}

\bigskip

\noindent
An interesting and profound idea about the degrees of freedom
describing gravity, is the holographic principle
\cite{Hooft,Susskind}. 

\bigskip

\noindent
For the case of a sphere, with mass $M$,  of radius $r$, where $r$ will be the co-moving
radius for the expanding universe, a form, of
the holographic principle, states
that the ration $\epsilon = r_S/r$, has an upper limit equal to that of a 
black hole, {\it i.e.}

\bigskip

\begin{equation}
\epsilon = \left( \frac{r_S}{r} \right) \leq 1.
\label{HP}
\end{equation}

\bigskip

\noindent
With $G$ as Newton's constant, $r_S = 2 G M$ 
is the Schwarzschild radius.
It is interesting,
from the viewpoint of the physical understanding of the
visible universe, to use accurate observational data to
check, whether the simplied, and non-covariant, Eq.(\ref{HP})
is satisfied at the present time, $t=t_0$, and in the past,
cognizant that, with dark energy,
if $r$ sufficiently increases, Eq.(\ref{HP})
will eventually be violated, if the
universe expands exponentially quickly, and if one
assumes the approximations used here. Note that, at the time
of \cite{Hooft}, before
dark energy, if Eq.(\ref{HP}) is now satisfied, 
one might expect it to remain so.

\bigskip

\noindent
The holographic principle is supported, by string theory. The AdS/CFT correspondence
\cite{AdSCFT} is an explicit realization of Eq.(\ref{HP}), and so, apart from the non-trivial subtlety that our universe is dS, not AdS, from the viewpoint
of string theory, there is every reason to believe the 
holographic principle, and to wish to check Eq.(\ref{HP}).
It is related to recent considerations of the entropy of the universe
\cite{Verlinde,EFS1,EFS2}.

\bigskip

\noindent
However, physics is an empirical science, and therefore the scientific method
dictates that we should find a physical example, in which Eq.(\ref{HP})
can be calculated. The result, reported here, is that a detailed and accurate check of Eq.(\ref{HP}), as applied to
the visible universe, fails, by a statistically-significant amount, although
in the past, a few billion years ago, it was satisfied.

\bigskip

\noindent
I should define, precisely, what is meant by the visible universe. It is the
sphere centered for convenience at the Earth with radius
$d_A(Z^*) = 14.0 \pm 0.1 Gpc$. The value of $d_A(Z^*)$ is the particle horizon corresponding
to the recombination red shift $Z^* = 1090 \pm 1$, and is measured directly by WMAP7 \cite{WMAP7},
without needing the details of the expansion history. Thus, ``visible" means
with respect to electromagnetic radiation.

\section{The Visible Universe}

\bigskip

\noindent
The motion is that the visible universe, so defined, is a physical object
which should be subject to the holographic principle. It is an expanding,
rather than a static, object, yet my understanding is that the principle,
at least in its covariant form,
is still expected to be valid.

\bigskip

\noindent 
I shall use the notation employed by the WMAP7 paper \cite{WMAP7},
from which all observational data are taken. 

\bigskip

\noindent
The present age, $t_0$, of the universe
is measured to be

\begin{equation}
t_0 = 13.75 \pm 0.13 Gy
\label{age}
\end{equation} 

\bigskip

\noindent
The comoving radius, $d_A(Z^*)$,  of the visible universe, is,
likewise, measured to one percent accuracy:

\bigskip

\begin{equation}
d_A(Z^*) \equiv (1 + Z^*) D_A(Z^*) = c\int^{t_0}_{t^*} \frac{dt}{a(t)} =14.0 \pm 0.1 Gpc,
\label{d}
\end{equation}

\bigskip

\noindent 
where it is noted that the measurement, of $d_A(Z^*)$,
does not require knowledge, of the expansion history, $a(t)$, for $t^* \leq t \leq t_0$.

\bigskip

\noindent
The critical density, $\rho_c$, is provided by the formula

\bigskip

\begin{equation}
\rho_c = \left( \frac{3 H_0^2}{8 \pi G} \right),
\label{critical}
\end{equation}

\bigskip

\noindent
whose value depends  on $H_0$, as does the total, baryonic plus dark,
matter density, $\rho_m$

\bigskip

\begin{equation}
\rho_m \equiv \Omega_m \rho_c.
\label{matter}
\end{equation}

\bigskip

\noindent
Because the error on the Hubble parameter, $H_0$,
is several per cent, it is best to avoid $H_0$, in checking the holographic principle.

\bigskip

\noindent
The mass of the matter, M(Z*), contained in the visible universe, is
first estimated, to be augmented later,  as

\bigskip

\begin{equation}
M(Z^*) = \frac{4 \pi}{3} d_A(Z^*)^3 \rho_m
\label{M}
\end{equation}

\bigskip

\noindent
which gives $M(Z^*) \sim 5.5 \times 10^{23} M_{\odot}$,
consistent with the oft-used estimates of $\sim10^{11}$
galaxies, each of mass $\sim 10^{12} M_{\odot}$ including dark matter.
The Schwarzschild radius, $r_S(Z^*)$, is given by
$R_S(Z^*) \equiv 2 G M(Z^*)$.

\bigskip

\noindent
Collecting results enables the desired accurate check of the 
simplified holographic principle,
which compares the radius, $r$, for the visible universe 
with that for a black hole, $r_S$ of the same mass. 

\bigskip

\noindent
A shift parameter, $\sigma$, was defined by Bond, Efstathiou and Tegmark (BET)
in \cite{Bond}, as

\bigskip

\begin{equation}
\sigma = \frac{\sqrt{\Omega_m H_0^2}}{c} (1+Z^*) D_A(Z^*),
\label{shift}
\end{equation}

\bigskip

\noindent
which was, with great prescience, introduced by BET, as a dimensionless
quantity, to be measured, accurately, by CMB observations.

\bigskip

\noindent
This BET shift parameter, $\sigma$, of Eq. (\ref{shift}), is given in \cite{WMAP7},
as (Note that $\sigma$ is designated $R$ in \cite{WMAP7})

\bigskip

\begin{equation}
\sigma = 1.725 \pm 0.018.
\label{R}
\end{equation}

\bigskip

\noindent
A little algebra shows that the BET shift parameter $\sigma$ 
provides the most accurate 
available check, of the holographic principle, by virtue of the result

\bigskip

\begin{equation}
\epsilon (t_0) = \left[ \frac{r_S}{r} \right]  \equiv \sigma^2  = 2.976 \pm 0.062 
\label{HPcheck}
\end{equation}

\bigskip

\noindent
which is surprisingly close to the saturation
of Eq.(\ref{HP}). The derivation
of Eq.(\ref{HPcheck}) is as follows:

\bigskip

\begin{eqnarray}
\epsilon & = &\left( \frac{r_S}{r} \right)  \\
& = & \left( \frac{1}{(1 + Z^*) D_A (Z^*)} \right) \left( 2G. 
\frac{4 \pi}{3} (1 + Z^*)^3 D_A(Z^*)^3 \Omega_m \frac{3 H_0^2}{8 \pi G} \right) \\
& = & \Omega_m H_0^2 (1 + Z^*)^2 D_A(Z^*)^2 \\
& = & \sigma^2 = 2.976 \pm 0.062 ~~~~~~~~~   Q.E.D.
\end{eqnarray}

\bigskip

\noindent
I used the definition in Eq.(\ref{shift}) for this calculation. It is very fortunate that
the square of $\sigma$, defined in Eq. (\ref{shift}), agrees with
the calcualtion of $\epsilon$, in Eq. (\ref{HPcheck}).

\bigskip

\noindent
This is suggestive that the expression, Eq.(\ref{M}), overestimates
somewhat the mass because the density is smaller at the sphere's center, 
where the observer is.
More accurately, one can be to take into account this variation of density
with redshift by using instead

\bigskip

\begin{eqnarray}
M(Z^*) & = & \int_{0}^{d_A(Z^*)} dr 4 \pi r^2 \rho_m \left(\frac{r}{d_A(Z^*)}
\right)^3  \nonumber \\
& = & \frac{1}{2} \frac{4 \pi d_A(Z^*)^3 \rho_m}{3},
\label{Mnew}
\end{eqnarray}

\bigskip

\noindent
which goves $M(Z^*)m \sim 2.7 \times 10^{23} M_{\odot}$ and 
reduces the estimate of $\epsilon (t_0)$ by a factor two to 

\begin{equation}
\epsilon (t_0) = 1.48 \pm 0.03,
\label{HPrecheck}
\end{equation}

\bigskip

\noindent
with a discrepancy of only 48 percent which is
surprisingly small compared to the
size of either of the radii, both of which are $\sim 10^{61}$ in natural (Planck)
units. The mass of the visible universe in Eq.(\ref{Mnew}) for the volume in
asymptotically-flat space is a phenomenological estimate, but the best available.

\bigskip

\section{Discussion}

\noindent
To my knowledge, the visible universe is, at present, the only physical object,
for which it is possible to calculate, and compare with experiment, or observation,
the simplified holographic principle.

\bigskip

\noindent
The result Eq. (\ref{HPrecheck}) is surprisingly close to $\epsilon = 1$. The
departure from Eq.(\ref{HP}) can presumably be ascribed to the
two assumptions of (i) asymptotic flatness and (ii) a quasi-static
universe. It is therefore a very interesting challenge
to relax these assumptions and further to use
the precision WMAP7 data to attempt to confirm that the
holographic principle, Eq. (\ref{HP}),
is precisly saturated by the observable universe.

\bigskip

\noindent
Such a calculational check is under way. For example, one may try
replacing the Schwarzschild metric
by the McVittie metric\cite{McVittie} which accommodates both a cosmological constant
and an expanding universe. Whether this can confirm even more accurately the
holographic principle, only time will tell.

\bigskip

\section*{Acknowledgements}

\noindent
I am grateful to Professor Dam Son, for visiting
the University of Tokyo.  The idea, for how to address the dark energy
problem, occurred during Son's three lectures, about the holographic
principle, at Hongo campus, on February 6, 2010.
Useful discussions with Damien Easson, Hirosi Ooguri,
George Smoot and Shigeki Sugimoto are 
acknowledged. This work was supported in part by
the World Premier International Research Center Initiative 
(WPI initiative), MEXT, Japan and by U.S. 
Department of Energy Grant No. DE-FG02-05ER41418.

\end{document}